\documentclass[intlimits,twoside,a4paper]{article}

\usepackage[cp1251]{inputenc}

\usepackage[eqsecnum]{cmpj3}

\usepackage{bm}

\issue{2018}{21}{4}{43701}
\doinumber{10.5488/CMP.21.43701}

\title[Analytical method for calculation of the potential profiles of nitride-based resonance tunneling structures ]%
{Analytical method for calculation of the potential profiles of nitride-based resonance tunneling structures%
}%

\author[I.V. Boyko]{I.V. Boyko}
\address{Ternopil Ivan Puluj National Technical University, 56 Ruska St., 46001 Ternopil, Ukraine}

\date{Received May 8, 2018, in final form July 2, 2018}

\DeclareMathOperator{\Ai}{Ai}
\DeclareMathOperator{\Bi}{Bi}

\begin{document}

\maketitle

\begin{abstract}

Using the effective mass model for an electron and the dielectric continuum model, analytical solutions of the self-consistent Schr\"odinger-Poisson system of equations are obtained.
Quantum mechanical theory of electronic stationary states, the oscillator strengths of quantum transitions and a method of potential profile calculation is developed for the experimentally constructed three-well resonance-tunneling structure --- a separate cascade of quantum cascade detector.
For the proposed method, a comparison with the results of other methods and with the results of the experiment was carried out. A good agreement between the calculated value of the detected energy and its experimental value has been obtained, the difference being no more than $2.5\%$.
\pacs 73.21.Ac, 78.20.hb, 78.67.-n
\end{abstract}

\section{Introduction}

Quantum cascade lasers (QCL) \cite{1,2} and detectors (QCD) \cite{3,4,5,6,7,8} created experimentally on the basis of binary and ternary nitride alloys of ${\rm InN,}\, \, {\rm GaN,}\, \, {\rm AlN}$, etc., are now of considerable practical and theoretical interest. The physical properties of Group III-nitrides, their high temperature stability, large bandgap, and significant optical activity in particular, allow nanodevices created on their basis to operate efficiently in the actual infrared range of electromagnetic waves. Besides, an effective work of nitride QCL and QCD is possible within the range from cryogenic to room temperatures, which is a significant advantage in comparison with the nanodevices created on the basis of arsenide semiconductor compounds of GaAs, InAs, AlAs, which, in fact, can work only at cryogenic temperatures.

The anisotropy of  physical properties of nitride semiconductor materials is caused by strong interatomic bonds and by the fact that their crystal lattice is of the wurtzite type  hexagonal structure. Spontaneous and piezoelectric polarization arising in the layers of nitride semiconductor nanosystems is the result of uncompensated total dipole moment of crystal lattices and inconsistency of constant lattices at medium interfaces. As a result, internal electric fields appear in the nitride multilayer nanostructures causing a significant deformation of their potential profile.

The development of the consistent theory of electronic states and a general method for calculating the potential profiles of plane nitride-based resonance tunneling structures (RTS) is still an unsolved theoretical problem. The theory, which makes it possible to calculate the internal fields arising in the nitride semiconductor nanosystems, was developed in the papers \cite{9,10,11,12,13,14,15,16,17}. The application of this theory to the plane multilayer RTS is not a difficult problem, though it is not sufficient for constructing a method for calculating the RTS potential profiles, since it, for example, does not make it  possible to take into account the contribution of charge carriers to the value of the effective potential. Available methods for the calculation of RTS potential profiles have significant shortcomings both in terms of the completeness of the description of the effective potential components and in terms of the theoretical approaches used. One of the first methods was the one proposed in the paper \cite{18}. Since it takes into account only the contribution of internal electric fields in the effective potential for an electron, as it will be shown below, it can be used only for the qualitative estimations. In addition, widely used is the method based on the numerical solution of the Schr\"odinger and Poisson equations \cite{19,20,21}, as well as on the software \cite{22} which is often employed by experimenters for calculations \cite{4,6,23,24} based on the $k\cdot p$  method. Generally, the disadvantages of most numerical methods for RTS potential profiles calculation include the linearization of the Schr\"odinger and Poisson equations at the initial stage of their solution, as well as the neglect of the influence of the boundary conditions for the wave function and the fluxes of its probability. Besides, the application of such methods considerably complicates  the possibility of a further investigation of electronic transitions and electronic tunnel transport, electron-phonon interaction in such RTS and other important theoretical problems.

In the presented paper, analytical solutions of the self-consistent Schr\"odinger-Poisson system of equations are obtained. Using these solutions, the theory of stationary electronic states and oscillator strengths of quantum electronic transitions is developed for the three-well cascade of an experimentally realized QCD of near-infrared spectral region \cite{5}. A method for calculating the potential RTS profile is proposed. For the proposed method, the results are compared with the experimental results and with the results obtained by other methods. By calculating the energy spectrum of the electron and the oscillator strengths of quantum transitions, depending on the relative position of the active band and the extractor, geometric configurations of the cascade ensuring its effective operation as an active element of QCD were established.

\smallskip

\section{Statement of the problem. Components of the effective potential of nitride-based resonance tunneling structure }

\begin{figure}[!b]
\centerline{\includegraphics[width=0.65\textwidth]{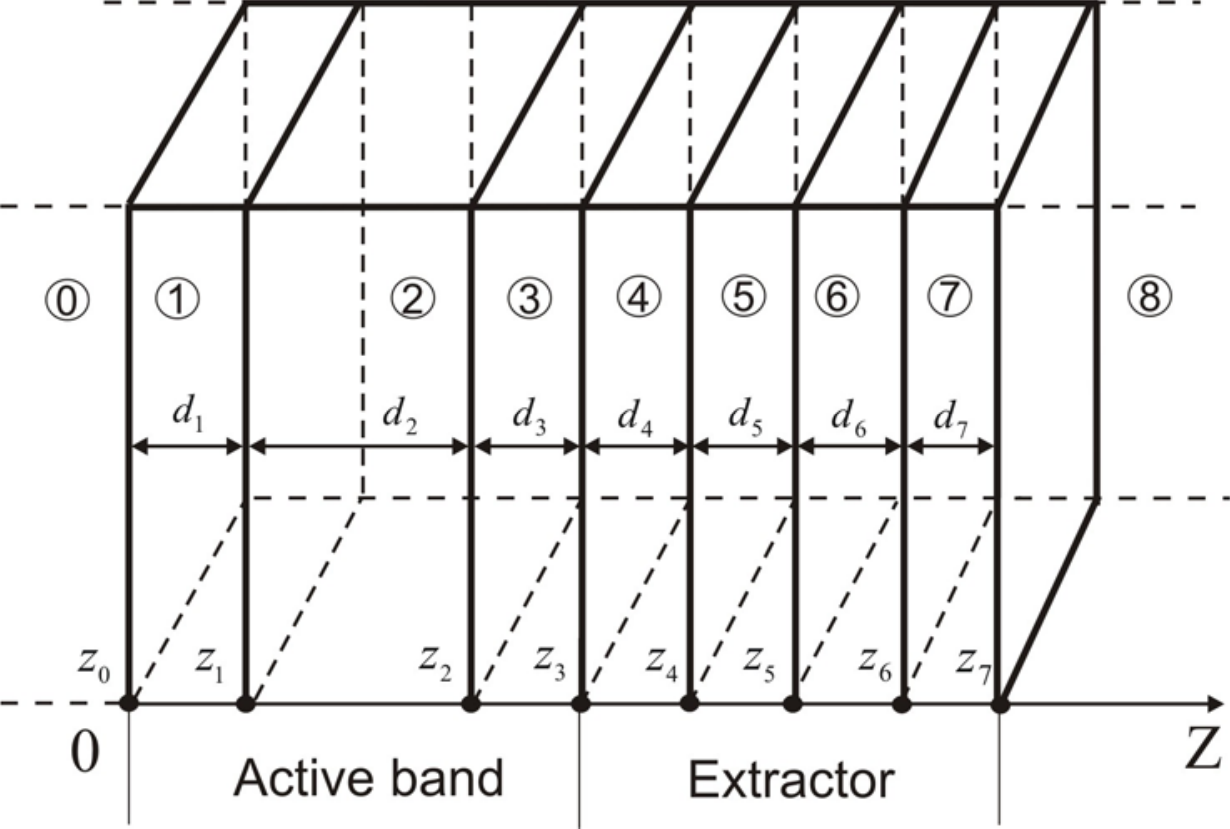}}
\caption{Geometrical scheme of the nitride-based RTS. $d_{1},d_{2},\ldots,d_{7}$  are RTS layers thickness, $z_{0},z_{1},\ldots,z_{7} $  are the boundary coordinates between these layers.} \label{fig1}
\end{figure}

In the Cartesian coordinate system, the three-well RTS--the cascade of QCD which consists of an active band and an extractor is considered (figure~\ref{fig1}). The coordinate system is chosen in such a way that its axis $OZ$ is perpendicular to the separation boundaries of the nanosystem layers. In accordance with the experimental work \cite{5}, it is assumed that the media (0) and (8), to the left and right of the RTS, correspond to the unstressed $\rm AlN$ medium, medium (2) corresponds to the semiconductor material $\rm GaN$, media (4), (6) --- $\rm Al_{0.25}Ga_{0.75}N$ semiconductor material, medium (1), (3), (5), (7) --- $\rm AlN$ semiconductor material.

Using the effective mass model and the dielectric continuum model, the effective electron mass and dielectric permeability of the RTS can be written as:
\begin{align}
m(z)&=m_{0} [\theta (z-z_{1} )+\theta (z-z_{2} )]+m_{1} \left\{\theta (-z)+\theta (z-z_{7} )+\sum _{p=0}^{3}\left[\theta (z-z_{2p} )-\theta (z-z_{2p+1} )\right] \right\} \nonumber\\  &+m_{2} \sum _{p=2}^{3}\left[\theta (z-z_{2p-1} )-\theta (z-z_{2p} )\right] ,\\
\varepsilon (z)&=\varepsilon _{(0)} \left\{\theta (-z)+\theta (z-z_{7} )+\sum _{p=0}^{3}\left[\theta (z-z_{2p} )-\theta (z-z_{2p+1} )\right] \right\}+\varepsilon _{(1)} [\theta (z-z_{1} )+\theta (z-z_{2} )] \nonumber\\   &+\varepsilon _{(2)} \sum _{p=2}^{3}\left[\theta (z-z_{2p-1} )-\theta (z-z_{2p} )\right] ,
\end{align}
where $\theta(z)$ is the Heaviside step function, $m_{0} =m^{(2)}$; $m_{1} =m^{(0)} =m^{(1)} =m^{(3)} =m^{(5)} =m^{(7)}= m^{(8)}$; $m_{2} =m^{(4)} =m^{(6)}$ is the effective electron mass in the potential barriers and wells of the RTS, $\varepsilon _{(0)} =\varepsilon ^{(0)} = \varepsilon ^{(1)} =\varepsilon ^{(3)} =\varepsilon ^{(5)} =\varepsilon ^{(7)} =\varepsilon ^{(8)}$; $\varepsilon _{(1)} =\varepsilon ^{(2)}$; $\varepsilon _{(2)} =\varepsilon ^{(4)} =\varepsilon ^{(6)}$  is dielectric permeability of nanostructure material layers, respectively.

The magnitude of the macroscopic polarization $P^{(p)}$  that arises in an arbitrary $p$-th RTS layer is the sum of  spontaneous $P_\text{Sp}^{(p)}$  and piezoelectric $P_\text{Pz}^{(p)}$  polarizations calculated according to the general theory \cite{9,10,11}:
\begin{align}
P^{(p)} =P_\text{Sp}^{(p)} +P_\text{Pz}^{(p)},
\end{align}
where the corresponding polarization $P_\text{Pz(Sp)}^{(p)}$ for a ternary semiconductor A$_{x}$B$_{1-x}$N,  depending on the concentration $x$ of the component A,  is determined in the linear approximation:
\begin{align}
P_\text{Pz(Sp)}^{(p)} (x)=P_\text{Pz(Sp)}^{\text{AN}\,(p)} (x)+(1-x)P_\text{Pz(Sp)}^{\text{BN}\,(p)} (x).
\end{align}

The internal electric fields ($F_{p}$, $p=1,\ldots,7$) are determined from the continuity condition of the electrical displacement vector ${D}_{p} =\varepsilon ^{(p)} {F}_{p} +{P}^{(p)}$  on all boundaries of the RTS layers \cite{18,25}:
\begin{align}
D_{p} =D_{p+1}\,, 
\label{2.5}
\end{align}
and also from the condition that is satisfied for the total value of the voltage applied to the RTS \cite{11}:
\begin{align}
\sum _{p=1}^{7}F_{p}  d_{p} =0.
\label{2.6}
\end{align}
Having solved the system of equations (\ref{2.5}) and (\ref{2.6}), we obtain the expression for the electric field in an arbitrary layer of the RTS \cite{18}:
\begin{align}
F_{p} =\mathop{\sum _{k=1}^{7}}\limits_{ (k\ne p)}\left[P^{(k)} -P^{(p)} \right]\frac{d_{k} }{\varepsilon ^{(k)} }   \Big/    \varepsilon ^{(p)} \sum _{k=1}^{7}\frac{d_{k} }{\varepsilon ^{(k)} }  \,,
\label{2.7}
\end{align}
$d_{k}$   is the thickness of the corresponding RTS layer.

According to the papers \cite{19,21,25,26}, the effective potential of the RTS for an electron will be determined as the sum of the components:
\begin{align}
V(z)=\Delta E_\text{C} (z)+V_\text{E} (z)+V_\text{H} (z)+V_\text{HL} (z).
\label{2.8}
\end{align}
In the expression (\ref{2.8}):
\begin{align}
\Delta E_\text{C} (z)=\left\{
\begin{array}{l} 
{0.765[E_{g} ({\rm AlN})-E_{g} ({\rm GaN})], \hspace{16.5mm} z<0,\, \, 0\leqslant z<z_{1}\, ,\, \, z_{2} \leqslant z<z_{3}\,,  } \\ 
{\hspace{58.2mm} z_{4} \leqslant z<z_{5}\,, \ \ z_{6} \leqslant z<z_{7} \,,\, \, z>z_{7}\,, } \\ 
{0,\hspace{55.2mm} z_{1} \leqslant z<z_{2}\,, } \\ 
{0.765[E_{g} ({\rm Al}_{0.25} {\rm Ga}_{0.75} {\rm N})-E_{g} ({\rm GaN})],\, \, \, \, z_{1} \leqslant z<z_{2} } 
\end{array}\right.
\label{2.9}
\end{align}
is the potential RTS profile for an electron, the calculation of which was performed without taking into account the electric field of the piezoelectric and spontaneous polarizations. The dependence of the bandgap of the  ${\rm Al}_{x} {\rm Ga}_{1-x} {\rm N}$ semiconductor on temperature $T$ in expression (\ref{2.9}) is calculated from the Varshni linear-quadratic relation \cite{27}:
\begin{align}
E_{g} (x,T)=E_{g} (x,0)-\frac{a(x)T^{2} }{b(x)+T}\,,
\label{2.10}
\end{align}
where the bandgap at zero temperature, depending on the magnitude $x$, can be represented as:
\begin{align}
E_{g} (x,0)=xE_{g} ({\rm AlN})+(1-x)E_{g} ({\rm GaN})+\alpha x(1-x).
\label{2.11}
\end{align}
In the expression (\ref{2.11}): $E_{g} ({\rm AlN})=6.25$~eV,  $E_{g} ({\rm GaN})=3.51$~eV  is the bandgap of $\rm AlN$  and $\rm GaN$  semiconductor correspondingly, $\alpha =0.7$~eV  is a bowing parameter for Group III-nitrides \cite{27}, $a(x)=[1.799x + 0.909(1-x)]\cdot 10^{ -3}$~(eV/K), $b(x)=1462x+830(1 -x)$~(K)  are the Varshni parameters \cite{27}.

The component $V_\text{E}(z)$  of the effective potential --- the potential energy characterizing the contribution of the electron interaction with the internal fields ($F_{p}$, $p=1,\ldots,7$) of spontaneous and piezoelectric polarizations arising in the RTS, is determined by the expression:
\begin{align}
V_\text{E} (z)=e\sum _{p=1}^{7}(-1)^{p-1} (F_{p} z-F_{p-1} z_{p-1} )\left[\theta (z-z_{p-1} )-\theta (z-z_{p} )\right] ,\qquad F_{0} =0.
\label{2.12}
\end{align}

The potential $V_\text{H}(z)$  is determined by the contribution of charge carriers within the RTS, and its immediate calculation will be carried out further.

Component:
\begin{align}
V_\text{HL} (z)=-\frac{1}{4\piup } \left(\frac{9}{4\piup ^{2} } \right)^{{1 \mathord{\left/{\vphantom{1 3}}\right.\kern-\nulldelimiterspace} 3} } \left[1+\frac{0.6213r_{s} }{21} \ln \left(1+\frac{21}{r_{s} (z)} \right)\right]\frac{e^{2} }{\varepsilon _{0} r_{s} (z)\varepsilon (z)a_\text{B}^{*} (z)}
\label{2.13}
\end{align}
is the Hedin-Lundquist exchange-correlation potential \cite{26,28}, where $r_{s} (z)=[4\piup a_\text{B}^{*3} n(z)/3]^{-1/3} $   is the dimensionless function, $a_\text{B}^{*} (z)=a_\text{B}\varepsilon (z)/m(z)$, $a_\text{B}$   is the Bohr radius, $n(z)$  is the concentration of carriers forming a static space charge.

\section{Method for determination of the solutions of the self-consistent\\ Schr\"odinger and Poisson equations system}

The stationary spectrum of the electron and its wave functions $\Psi(z)$  are determined by solutions of the self-consistent Schr\"odinger-Poisson system of equations:
\begin{align}
\left\{\begin{array}{l} \displaystyle{-\frac{\hbar ^{2} }{2} \frac{\rd}{\rd z} \left[\frac{1}{m(z)} \frac{\rd\Psi (z)}{\rd z} \right]+V(z)\Psi (z)=E\Psi (z),} \vspace{2mm} \\ 
\displaystyle{\frac{\rd}{\rd z} \left[\varepsilon (z)\frac{\rd V_\text{H} (z)}{\rd z} \right]=-e\rho (z)}, \end{array}\right.
\label{3.1}
\end{align}
$\rho(z)$  is the total charge density at an arbitrary point $z$ inside the RTS.

The method of the self-consistent system solutions determination (\ref{3.1}) is as follows. At first, keeping in the effective potential the most significant terms $\Delta E_\text{C}(z)$  and $V_\text{E}(z)$, the solutions of the Schr\"odinger equation must be found. The solutions of the Schr\"odinger equation are determinated in every  RTS region. Therefore, the wave function that takes into account its finiteness, can be presented as follows:
\begin{align}
\Psi _{0} (z)&=\Psi _{0}^{(0)} (z)\theta (-z)+\Psi _{0}^{(8)} (z)\theta (z-z_{7} )+\sum _{p=1}^{7}\Psi _{0}^{(p)} (z) \left[\theta (z-z_{p-1} )-\theta (z-z_{p} )\right]\nonumber \\ 
&=A_{0}^{(0)} \re^{\,\chi ^{(0)} z} \theta (-z)+\sum _{p=1}^{7}\left[A_{0}^{(p)} \Ai(\zeta ^{(p)} (z))+B_{0}^{(p)} \Bi(\zeta ^{(p)} (z))\right] \left[\theta (z-z_{p-1} )-\theta (z-z_{p} )\right] \nonumber\\ 
&+B_{0}^{(8)} \re^{-\chi ^{(8)} z} \theta (z-z_{7} ),
\end{align}
where $A_{0}^{(0)}$, $B_{0}^{(8)}$  are the coefficients in the solutions of the Schr\"odinger equation to the left and to the right of the RTS, respectively, and $A_{0}^{(p)}$, $B_{0}^{(p)}$  are inside the RTS, $\Ai(\zeta )$, $\Bi(\zeta )$  are the Airy functions,
\begin{align}
\zeta ^{(p)} (z)=\left[2m^{(p)} eF_{p} /\hbar ^{2} \right]^{1/ 3 } &\left\{[\Delta E_\text{C} (z)-E]/eF_{p} -z\right\}, \quad \chi ^{(0)} =\chi ^{(8)} =\hbar ^{-1} \sqrt{2m_{0} (U -E)}, \\ 
U&=0.765[E_{g} ({\rm AlN})-E_{g} ({\rm GaN})]. 
\end{align}
The lower index in the coefficients $A_{0}^{(0)}$, $B_{0}^{(8)}$, $A_{0}^{(p)}$, $B_{0}^{(p)}$  corresponds to the order of approximation.

From the continuity condition of the wave function and the fluxes of its probability density on all boundaries of the nanosystem:
\begin{align}
\Psi ^{(p)} (z_{p} )=\Psi ^{(p+1)} (z_{p} ); \qquad  \left.\frac{\rd\Psi^{(p)} (z)}{m(z)\rd z} \right|_{z=z_{p} -\varepsilon } =\left. \frac{\rd\Psi^{(p+1)} (z)}{m(z)\rd z} \right|_{z=z_{p} +\varepsilon } ,
\label{3.5}
\end{align}
a dispersion equation from which the energy spectrum of the electrons ($E_{n}$)  is obtained as well as all coefficients $A_{0}^{(0)}$, $B_{0}^{(8)}$, $A_{0}^{(p)}$, $B_{0}^{(p)}$   expressed through one of them are found. The last coefficient is determined from the condition of the wave function normalization
\begin{align}
\int _{-\infty }^{+\infty }\Psi _{n}^{*} (E_{n} ,z)\Psi _{n} (E_{n} ,z) \rd z=1,
\label{3.6}
\end{align}
which uniquely determines the wave functions of all the stationary states of the electron.
The charge density within the RTS is found as follows:
\begin{align}
\rho (z)=e[N_\text{D}^{+} -n(z)]+\sum _{p=1}^{7}\sigma _{p} \delta (z-z_{p} ),
\end{align}
where the concentration of ionized donor impurities is given by
\begin{align}
N_\text{D}^{+} = \frac{N_\text{D} }{1+g\exp  \left(\frac{E_\text{F} -E_{n} }{k_\text{B}T} \right)}\,,
\end{align}
where $N_\text{D}$  is the concentration of donor impurities, $g=2$ is the degeneracy factor,
\begin{align}
n(z)=n_{0} (z)\sum _{n}\left|\Psi (E_{n} , z)\right|^{2}  \ln \left|1+\exp \left(\frac{E_\text{F} -E_{n} }{k_\text{B} T} \right)\right|,\qquad n_{0} (z)=\frac{\, m(z)k_\text{B} T}{\piup \hbar ^{2} }
\end{align}
is the electron concentration in the nanostructure,
\begin{align}
\sigma _{p} =P^{(p+1)} -P^{(p)}
\end{align}
is the surface density of the charge carriers arising at the boundaries of the RTS layers due to different values of the total polarization.
Now, in the region of an arbitrary layer of the RTS, the Poisson equation can be written as follows:
\begin{align}
\frac{\rd^{2} V_\text{H}^{(p)} (z)}{\rd z^{2} } =-\frac{e}{\varepsilon ^{(p)} } \left\{e\left[N_\text{D}^{+} -\frac{m^{(p)} k_\text{B} T}{\piup \hbar ^{2} } \sum _{n}\left|\Psi (E_{n} , z)\right|^{2}  \ln \left|1+\exp \left(\frac{E_\text{F} -E_{n} }{k_\text{B} T} \right)\right|\right]+\sigma _{p} \delta (z-z_{p} )\right\}
\end{align}
and its exact analytic solution that takes into account the fact that for stationary electronic states $\Psi (E_{n} , z)=\Psi ^{*} (E_{n} , z)$  is given by
\begin{align}
V_\text{H}^{(p)} (z)&=-\frac{e}{\varepsilon ^{(p)} } \int _{0}^{z}\int _{0}^{x_{1} }\left\{e\left[N_\text{D}^{+} -\frac{ k_\text{B} T}{\piup \hbar ^{2} } \sum _{n}\left|\Psi (E_{n} , x_{2} )\right|^{2}  \ln \left|1+\exp \left(\frac{E_\text{F} -E_{n} }{k_\text{B} T} \right)\right|\right]+ \sigma _{p} \delta (x_{2} -z_{p} )\right\}\rd x_{1} \rd x_{2}   
\nonumber \\ 
&=-\frac{e}{\varepsilon ^{(p)} } \left[\frac{eN_\text{D}^{+} }{2} (z-z_{p-1} )^{2} +(z-z_{p} )\sigma _{p} \theta (z-z_{p} )\right]
\nonumber \\ 
&+\frac{1}{3\piup } \left[\frac{e m^{(p)}}{4\hbar ^{4} F_{p}^{2} } \right]^{{1/3} } \frac{k_\text{B} T}{\varepsilon ^{(p)} } \sum _{n}\ln \left|1+\exp \left(\frac{E_\text{F} -E_{n} }{k_\text{B} T} \right)\right| \bigg(\left\{2[A_{0}^{(p)}  ]^{2} \right. [\zeta ^{(p)} (z)]^{2}\Ai^{2}(\zeta ^{(p)} (z)) 
\nonumber \\
&\left. -\Ai(\zeta ^{(p)} (z))\Ai'(\zeta ^{(p)} (z))-2\zeta ^{(p)} (z)\Ai'^{2}(\zeta ^{(p)} (z)) \right\}+\left\{2[B_{0}^{(p)} \right. ]^{2} [\zeta ^{(p)} (z)]^{2}\Bi^{2}(\zeta ^{(p)} (z)) 
\nonumber \\ 
&\left. -\Bi(\zeta ^{(p)} (z))\Bi' (\zeta ^{(p)} (z))-2\zeta ^{(p)} (z)\Bi'^{2} (\zeta ^{(p)} (z)) \right\}+\left\{A_{0}^{(p)} B_{0}^{(p)} \Ai(\zeta ^{(p)} (z)) \right.  \nonumber \\ 
&\times\big\{4[\zeta ^{(p)} (z)]^{2} \Bi(\zeta ^{(p)} (z)) 
\left. -\Bi'(\zeta ^{(p)} (z))\big\}+\Ai'(\zeta ^{(p)} (z))\big[\Bi(\zeta ^{(p)} (z))+4\zeta ^{(p)} (z)\Bi'(\zeta ^{(p)} (z))\big]\right\}\bigg) \nonumber \\
&+C_{1}^{(p)} (z-z_{p-1} )+C_{2}^{(p)} .
\label{3.12}
\end{align}

The coefficients $C_{1}^{(p)}$, $C_{2}^{(p)}$  are uniquely determined from the boundary conditions of the continuity of the potential $V_\text{H} (z)$  and from the corresponding electric displacement vector at all the RTS boundaries:
\begin{align}
V_\text{H}^{(p)} (z_{p} )=V_\text{H}^{(p+1)} (z_{p} ); \qquad &\left.\varepsilon ^{(p)}\frac{ \rd V_\text{H}^{(p)} (z)}{\rd z} \right|_{z=z_{p} -\varepsilon } -\left. \varepsilon ^{(p+1)}\frac{ \rd V_\text{H}^{(p+1)} (z)}{\rd z} \right|_{z=z_{p} +\varepsilon } =-\sigma (z_{p} ); \nonumber\\ 
&\varepsilon \to +0;\qquad p=0,\ldots,7
\label{3.13}
\end{align}
as well as from the conditions for the disappearance of potential $V_\text{H} (z)$ outside the RTS:
\begin{align}
\left. V_\text{H} (z)\right|_{z\to 0} \to 0;\qquad \left. V_\text{H} (z)\right|_{z\to z_{7} } \to 0.
\end{align}
Then, the potential  $V_\text{H} (z)$ can be presented as:
\begin{align}
V_\text{H} (z)=\sum _{p=1}^{7}V_\text{H}^{(p)} (z) \left[\theta (z-z_{p-1} )-\theta (z-z_{p} )\right].
\end{align}

Further, for the total effective RTS potential $U_{{\rm eff}} (z)=V(z)$, calculated according to relations (\ref{2.8}), (\ref{2.9}), (\ref{2.12}), (\ref{2.13}), (\ref{3.12}), its linearization is performed. Then, the approximated effective potential of the RTS for an electron looks as follows:
\begin{align}
\tilde{U}_{{\rm eff}} (z)=\sum _{p=1}^{7}\sum _{l=0}^{N}eF(z_{p_{l} } )  z\left[\theta (z-z_{p_{l} } )-\theta (z-z_{p_{l+1} } )\right],
\label{3.16}
\end{align}
where
\begin{align}
F(z_{p_{l} } )=\frac{V(z_{p_{l+1} } )-V(z_{p_{l} } )}{e(z_{p_{l+1} } -z_{p_{l} } )}
\label{3.17}
\end{align}
is the effective value of the  internal electric field magnitude in the RTS,
\begin{align}
z_{p_{l} } =\frac{l}{2N} (z_{p} -z_{p-1} ),\qquad p=1,\ldots,7,\qquad z_{1_{0}} =0,
\label{3.18}
\end{align}
$N$  is the number of partitions in the  $p$-th nanosystem layer.

Having substituted the effective potential of the RTS in the form (\ref{3.16}) into the Schr\"odinger equation in the system (\ref{3.1}), taking into account (\ref{3.17}) and (\ref{3.18}), we obtain its solution
\begin{align}
\Psi _{0} (z)&=A_{1}^{(0)} \re^{\,\chi ^{(0)} z} \theta (-z)+B_{1}^{(8)} \re^{-\chi ^{(8)} z} \theta (z-z_{7} ) \nonumber\\ 
&+\sum _{p=1}^{7}\sum _{l=0}^{N}\left[A_{1}^{(p_{l} )} \Ai(\zeta ^{(p_{l} )} (z))+B_{1}^{(p_{l} )} \Bi(\zeta ^{(p_{l} )} (z))\right] \left[\theta (z-z_{p_{l} } )-\theta (z-z_{p_{l+1} } )\right],
\label{3.19}
\end{align}
where
\begin{align}
\zeta ^{(p_{l} )} (z)=\left[2m^{(p_{l} )} eF(z_{p_{l} } )/\hbar ^{2} \right]^{{1 / 3} } \left\{[\Delta E_\text{C} (z)-E]/eF(z_{p_{l} } )-z\right\},
\end{align}
\begin{align}
m^{(p_{l} )} =\left\{\begin{array}{l} 
m_{0}\,,\ \ z_{1} \leqslant z<z_{2} ; \\ 
m_{1}\, ,\, \, z<0,\ \ 0\leqslant z<z_{1}\,, \ \ z_{2} \leqslant z<z_{3}\,, \, \, z_{4} \leqslant z<z_{5}\,, \, \,  z_{6}\leqslant z<z_{7}\,, \, \, z>z_{7} ; \\ 
m_{2} \,,\ \ z_{3} \leqslant z<z_{4}\,,\ \ z_{5}\leqslant z<z_{6}. \end{array}\right.
\end{align}

By substituting the relation (\ref{3.19}), an expression for the component   of effective potential, calculated in the first iteration order, is given by
\begin{align}
\begin{array}{l} \displaystyle { V_\text{H} (z)=\sum _{p=1}^{7}\sum _{l=0}^{N}V_\text{H}^{(p_{l} )} (z) \left[\theta (z-z_{p_{l} } )-\theta (z-z_{p_{l} +1} )\right]}, \end{array}
\label{3.22}
\end{align}
where the expression for $V_\text{H}^{(p_{l} )} (z) $ is obtained from the formula (\ref{3.12}) by replacing the index ($p\rightarrow p_{l}$) and the coefficients ($A_{0}^{(p)}\rightarrow A_{1}^{(p_{l})};\, B_{0}^{(p)}\rightarrow B_{1}^{(p_{l})}$), 
\[\sigma _{p_{l} } =\left\{\begin{array}{l} {\sigma _{p}\, ,\quad z_{p_{l} } =z_{p} \,,} \\ {0,\qquad z_{p_{l} } \ne z_{p}. } \end{array}\right.\]

The coefficients  $A_{1}^{(1)}$, $B_{1}^{(8)}$, $A_{1}^{(p_{l} )}$, $B_{1}^{(p_{l} )}$ are determined from boundary conditions similar to the conditions (\ref{3.5}) and the normalization condition (\ref{3.6}), and the coefficients $C_{1}^{(p_{l} )}$, $C_{2}^{(p_{l} )}$  --- from conditions similar to the boundary conditions (\ref{3.13}).

The iterative procedure described makes it possible to establish the self-consistent solutions of the Schr\"odinger and Poisson system of equations, as well as all the components of the effective potential of the RTS for an electron with the required accuracy, which is presented by the obvious relation:
\begin{align}
\delta =\frac{\left|n_{\nu +1} (z)-n_{\nu } (z)\right|}{n_{\nu } (z)}\,,
\label{3.23}
\end{align}
$\nu$  is number of approximation (iteration).

The wave function   defined in an arbitrary order of iterations [$\Psi (E,z)=\Psi ^{(\nu )} (E,z)$] makes it possible to calculate the oscillator strengths of quantum transitions within the formula
\begin{align}
f_{n,n' } =\frac{2(E_{n} -E_{n' } )}{\hbar ^{2} } \sum _{p=1}^{N}m_{p}  \left|\,\int _{z_{p-1} }^{z_{p} }z\Psi _{n}^{*(p)} (E_{n} ,z)\Psi _{n' }^{(p)} (E_{n' } ,z) \rd z\right|^{2}.
\end{align}

\subsection{Discussion of the results}

The calculation of the wave functions $\Psi_{n}(E_{n},z)$  and the energy spectrum $E_{n}$  of the stationary electronic states localized in the RTS, the oscillator strengths of quantum transitions $f_{n,n' }$, and the effective potential of the RTS $V(z)$  were performed using the theory mentioned above.

\begin{figure}[!b]
\centerline{\includegraphics[width=0.5\textwidth]{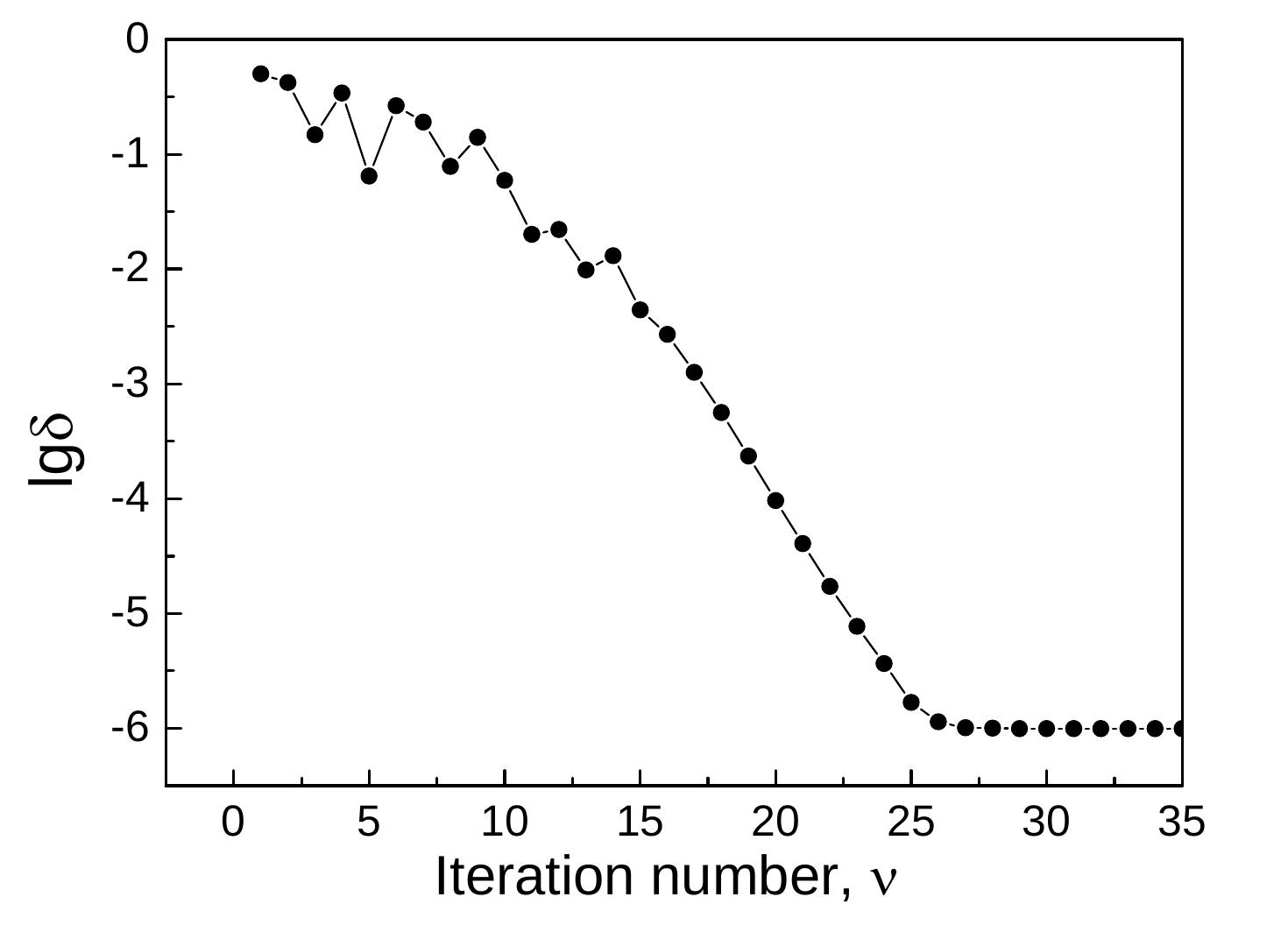}}
\caption{Dependence of calculation accuracy $\delta$ on the number of iterations $\nu$.} \label{fig2}
\end{figure}

Immediate numerical calculations were performed for the three-well RTS, i.e., the cascade of the experimentally realized QCD \cite{5}. The following geometric parameters of nanosystem were chosen: thickness of potential barriers $\Delta _{1} =\Delta _{2} =\Delta _{3} =\Delta _{4} =1.04$~nm, width of potential wells $d_{1} =1.56$~nm, $d_{2} =1.04$~nm, $d_{3} =1.04$~nm. The physical parameters of the RTS were taken from the papers \cite{27,29}. All calculations have been performed at temperature 300~K.

The convergence of the iterative procedure for finding solutions of the system of the Schr\"odinger and Poisson equations, whose accuracy of computation is described by the expression (\ref{3.23}), is presented in figure~\ref{fig2}, which shows the dependence of the logarithms of the accuracy of calculations $\delta$  on the number of iterations $\nu$. As it can be seen from the figure, the accuracy of calculations assumed in the presented paper is $\delta=10^{-6}$  is achieved by 23--25 iterations. Further iterations result in the output of the accuracy of the calculations on the saturation, which testifies the convergence of the iterative procedure of the proposed method.

\begin{figure}[!b]
\centerline{\includegraphics[width=0.5\textwidth]{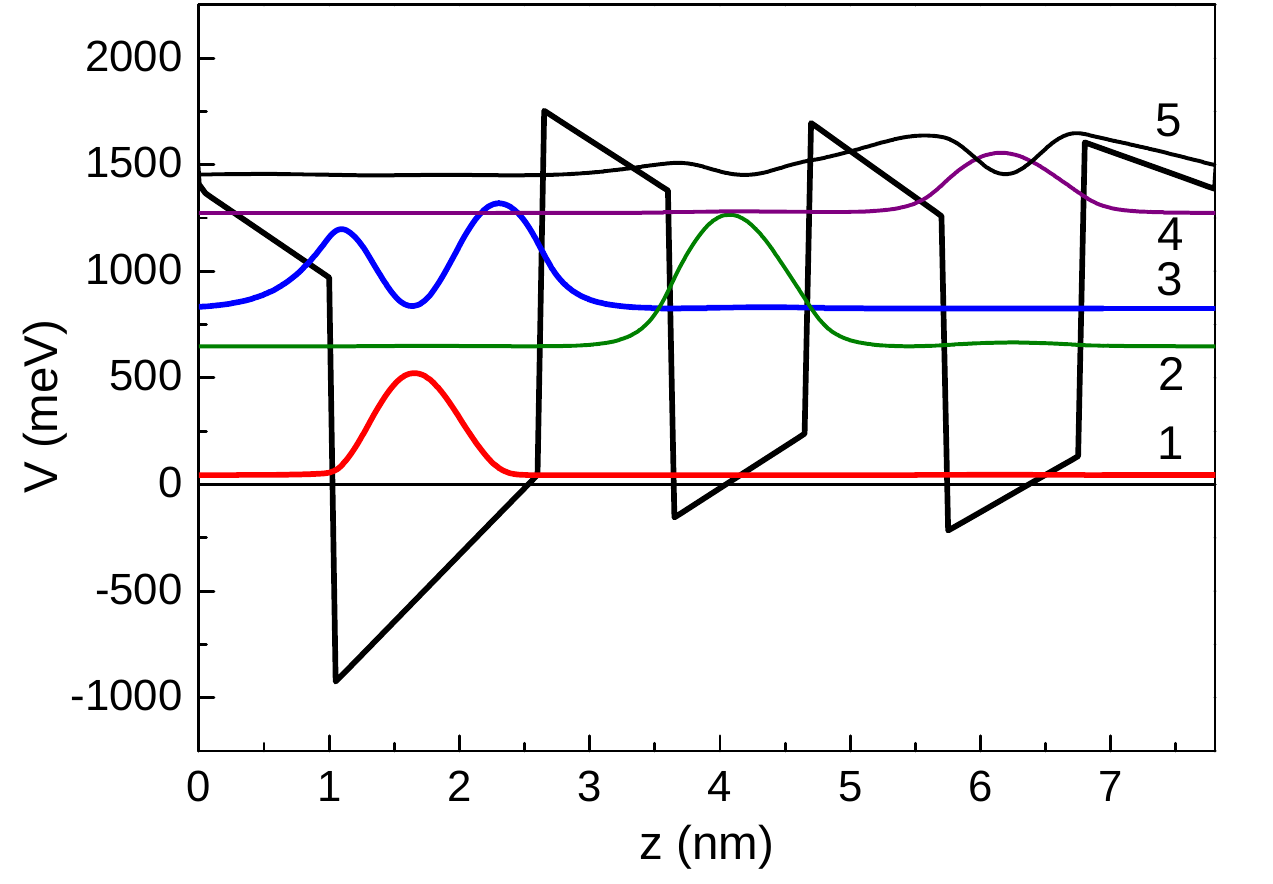}}
\caption{(Colour online) Potential profile of RTS and square moduli of the stationary states wave functions localized in the nanostructure. $n=1,2,\ldots,5$ are numbers of the corresponding energy levels. } \label{fig3}
\end{figure}

Figure~\ref{fig3} shows the energy scheme of the separate cascade of QCD, the calculation of the potential profile of which, depending on the magnitude, was performed according to the theory developed in the previous section according to the relations (\ref{2.9})--(\ref{2.13}), (\ref{3.22}). The square moduli of wave functions $\left|\Psi _{n} (E_{n} ,z)\right|^{2}$  for stationary electronic states localized in the RTS are also shown in the figure. It can be seen from the figure that the detector electronic transition is realized between the first and the third electronic states localized in the quantum well of the cascade active band.

\begin{figure}[!t]
\centerline{\includegraphics[width=0.99\textwidth]{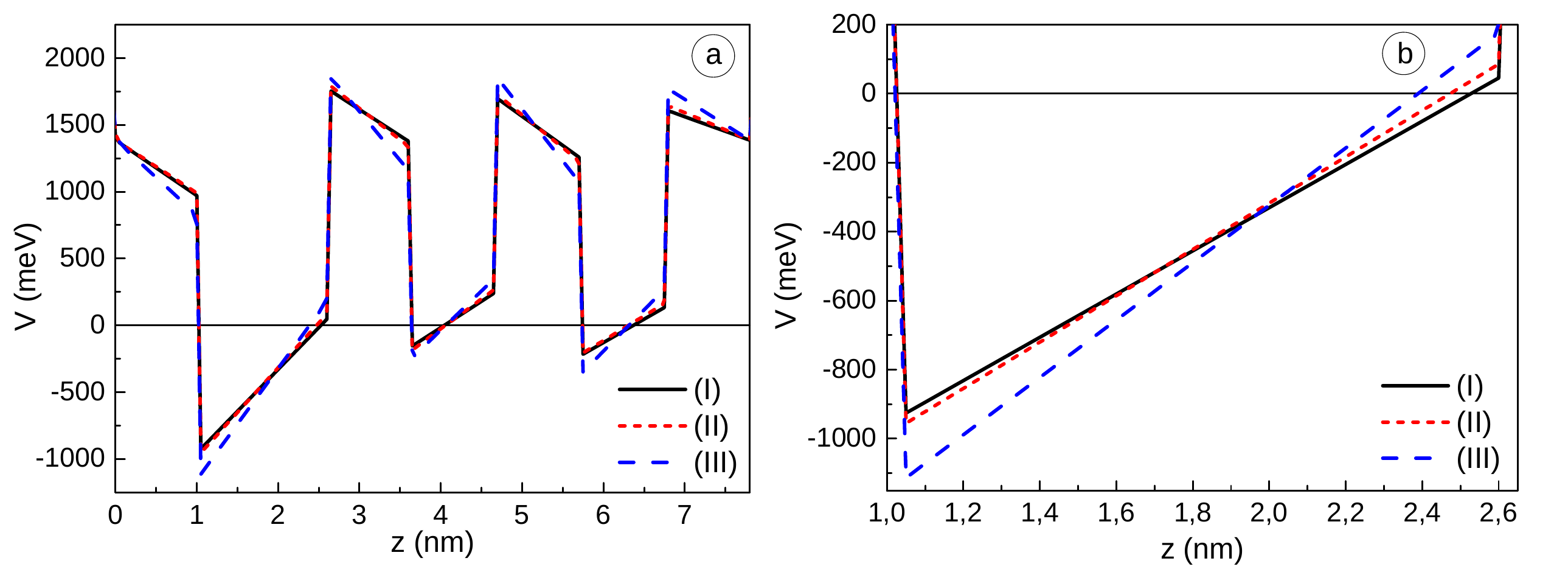}}
\caption{(Colour online) The dependence $V=V(z)$ of the effective potential RTS calculated in three models (a) and detailing  this dependence for the potential well of the active band (b).} \label{fig4}
\end{figure}

In order to  testify the effectiveness of the proposed method and to compare it with the results obtained by other methods, calculations of the potential RTS profile (figure~\ref{fig4}) were performed using the proposed method (I --- continuous line), using the method developed in \cite{25} (II --- short dashed line), and on the basis of the theory developed in \cite{18} (III --- dashed line). It can be seen from figure~\ref{fig4} that the dependences $V=V(z)$  obtained by all three methods are qualitatively identical. The dependences obtained by the methods (I) and (II) are very similar, and in dependence (III), the depths of potential wells and the height of potential barriers are larger than in dependences (I) and (II). This is due to the fact that in the theory developed in \cite{18}, only the contribution of internal fields arising in the nanosystem to the value of the effective potential is taken into account, which is determined from the expression (\ref{2.7}). Calculated in all three approaches, the values of the energies of electronic stationary states localized in the active zone of the cascade between which the detector transition occurs are as follows: $E_{1}^{({\rm I})} ={\rm 43.9}$~meV, $E_{3}^{({\rm I})} ={\rm 826.4}$~meV, $E_{1}^{({\rm II})} =41.4$~meV, $E_{3}^{({\rm II})} ={\rm 825.1}$~meV, $E_{1}^{({\rm III})} ={\rm -104.8}$~meV, $E_{3}^{({\rm III})} =802.2$~meV.

The value of the detected energy in the transition $1\rightarrow 3$ calculated on the basis of the developed theory $\Omega _{{\rm 13}}^{{\rm (I)}} ={\rm 782.5}$~meV, correlates well with the result obtained on the basis of the method (II) \cite{25} $\Omega _{{\rm 13}}^{{\rm (II)}} ={\rm 783.7}\, \, {\rm meV}$  and the experimental result $\Omega ^{({\rm exp})} ={\rm 8}00{\rm .}0\, {\rm meV}$,  the difference being  $2.5$\%. The result on the basis of the theory of paper \cite{18} gives the value of the transition energy $\Omega _{{\rm 13}}^{{\rm (III)}} ={\rm 9}07{\rm .0}\, \, {\rm meV}$. Besides, the values of the energies of the stationary electronic states are much smaller. Therefore, the application of such a theory is possible only for rough estimations.

An effective operation of the QCD is possible if the intensity of quantum transitions between the electronic states localized in the active zone of the cascade is the greatest. In this case, it is a transition $1\rightarrow3$  with energy $\Omega_{13}=E_{3}-E_{1}$. This means that, for a given geometric configuration of the studied RTS, the calculated oscillator strength of the quantum transitions between these states $f_{13}$  should be greater by an order of magnitude than the oscillator strengths for transitions from the first to the rest of the electronic states \cite{30,31}, that is:
\begin{align}
f_{13} >f_{1n' }\,, \quad n' =2,\ldots,5, \, \, n' \ne 3
\label{3.25}
\end{align}
as well as should exceed their sum:
\begin{align}
f_{13} >\sum _{n' =1}^{5}f_{1n' }\,, \quad n' \ne 3,
\label{3.26}
\end{align}
where the condition (\ref{3.26}) provides the detector reliable operation, at which the detection of other frequencies is excluded, which could result in QCD operating state failure.

\begin{figure}[!t]
\centerline{\includegraphics[width=0.5\textwidth]{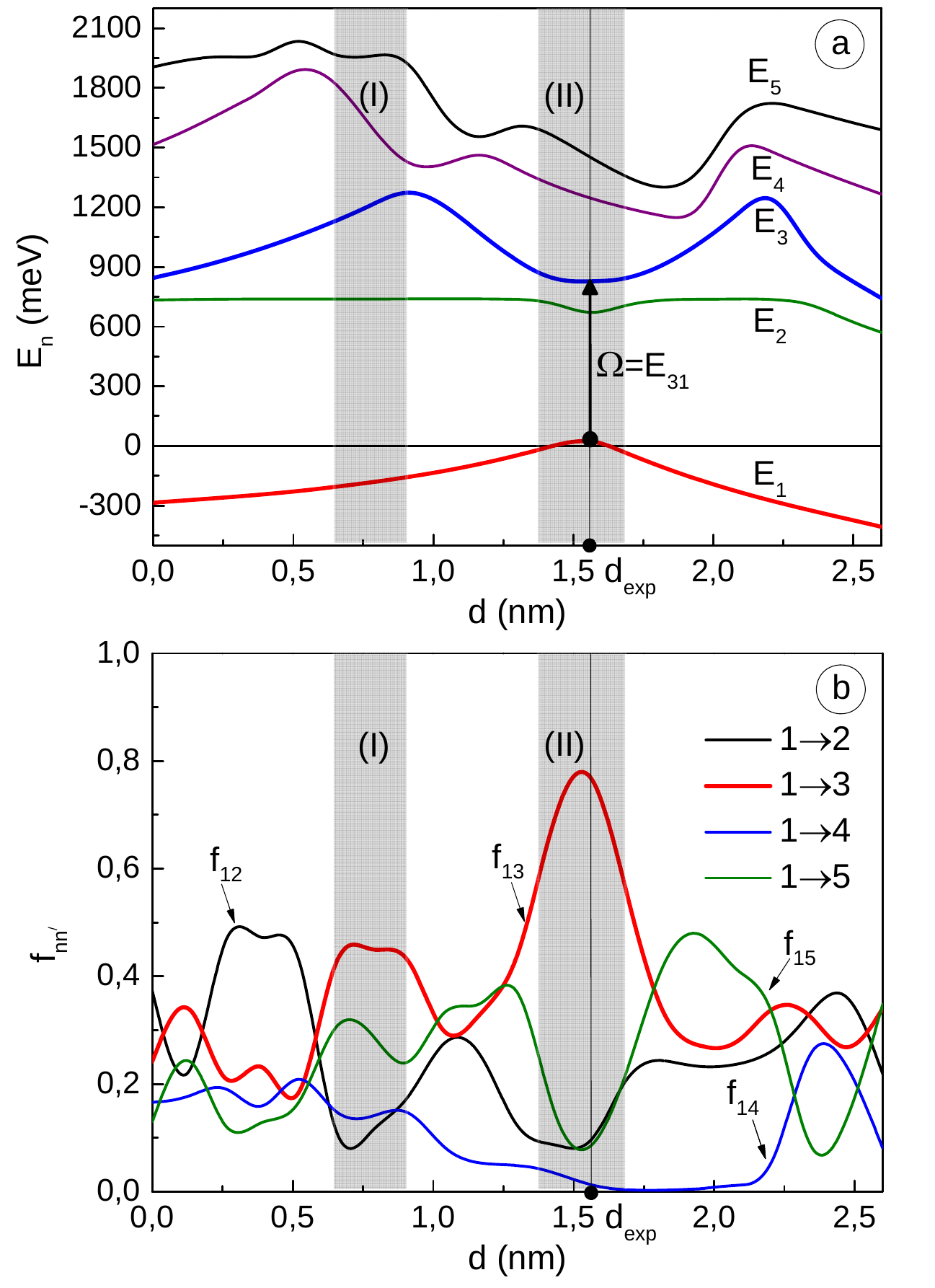}}
\caption{(Colour online) Dependences of the energies of the stationary electronic states ($E_{n}$, $n=1,2,\ldots,5$) (a) and the oscillator strengths of quantum transitions from the first stationary state to the remaining electronic states  ($f_{nn' }$) (b), calculated depending on the magnitude of $d=d_{1}+d_{3}$.} \label{fig5}
\end{figure}

In order to identify optimal configurations of the RTS [for which condition (\ref{3.25}) is satisfied], calculations of the stationary electron spectrum $E_{n}$  and the oscillator strengths of quantum transitions $f_{nn' }$  were performed as a function of the position  $d$ $(0\leqslant d\leqslant d_{1} +d_{3} )$ of the inner two-barrier structure [regions~(3)--(5) in figure~\ref{fig1}] relatively to the input [region~(1)] and to the output [region~(7)] of the cascade barriers with the unchanged remaining geometric parameters of the nanostructure. The results of these calculations are shown in figure~\ref{fig5}~(a), (b).

It is seen from figure~\ref{fig5}~(a) that the dependence $E_{n}=E_{n}(d)$ depends directly on $n$, and hence on the localization of the corresponding stationary electronic state. For example, the dependencies $E_{1}(d)$ and $E_{3}(d)$ whose states are localized in the input potential well (figure~\ref{fig3}), form one and two maxima, respectively. Similarly, in dependencies  $E_{4}(d)$ and $E_{5}(d)$, whose states  are localized in the output potential well, three and four maxima are formed, respectively. It should be noted that  $E_{2}(d)$ weakly depends on $d$, since this state is localized in the internal potential well of the RTS.

As it can be seen from figure~\ref{fig5}~(b), in the dependence $f_{nn' } =f_{nn' } (d)$, two intervals are formed, for which the magnitude of the oscillator strength $f_{13}$  is dominated by the remaining values of the calculated oscillator strengths. Condition~(\ref{3.25}) is satisfied for the first interval (I) ($0.65\, {\rm nm}\leqslant d\leqslant 0.91\, {\rm nm}$), but condition~(\ref{3.26}) is not fulfilled, since $f_{13}^{\text{(I)}} \approx f_{12}^{\text{(I)}} +f_{14}^{\text{(I)}} +f_{15}^{\text{(I)}}$. Therefore, for the considered geometric configuration of the cascade, the operation of the QCD will not be effective. For the second interval (II) ($1.38\, {\rm nm}\leqslant d\leqslant 1.69\, {\rm nm}$), which contains an experimentally realized configuration $d_{\exp } =1.56$~nm, both conditions~(\ref{3.25}) and (\ref{3.26}) are satisfied, and condition~(\ref{3.25}) is fulfilled more strictly: $f_{13}^{\text{(II)}} \gg f_{12}^{\text{(II)}} ,\, f_{14}^{\text{(II)}} ,\, f_{15}^{\text{(II)}}$. That is, for the geometric configuration $d_{\exp }$, we have: $f_{13}^{(\exp )} ={\rm 0.782}$; $f_{12}^{(\exp )} ={\rm 0.098}$; $f_{14}^{(\exp )} ={\rm 0.024}$; $f_{15}^{(\exp )} ={\rm 0.096}$. As it can be seen from figure~\ref{fig5}~(a), for the geometrical configuration found, the possibility of effective quantum detecting transitions with energies close to the experimentally obtained result 800~meV is realised.

Having summarised the obtained results, it can be concluded that the developed theory makes it possible to efficiently calculate the potential profiles of the nitride-based resonance-tunnel structures and to establish, using the proposed criteria, their geometric configurations providing an effective detection of the electromagnetic waves during operation of these nanosystems as active elements of the QCD.

\subsection{Conclusions}

1. The theory of stationary electronic states and the oscillator strengths of quantum transitions is developed on the basis of the found analytical solutions of the self-consistent Schr\"odinger-Poisson system of equations for three-well nitride RTS.

2. Using the developed theory, an analytical method for calculation of the nitride RTS potential profiles is proposed.

3. Direct calculations of the electron energy spectrum, of the oscillator strengths of quantum transitions and of the potential profile for the experimentally realized RTS, which functioned as an active element of QCD have been performed. Good reliability of the proposed theory in comparison with both numerical simulation results and experimental results is established.

\section*{Acknowledgements}\addcontentsline{toc}{section}{Acknowledgements}

The author is sincerely grateful and extends appreciation to the Head of the Chair of Theoretical Physics and Computer Modelling of Yu. Fed’kovych National University of Chernivtsi, Dr. Sci., Professor M.V. Tkach for his detailed comments on the obtained results and the content of the presented paper.

\ukrainianpart

\title{Аналітичний метод розрахунку потенціальних профілів нітридних резонансно-тунельних структур}

\author{І.В. Бойко}
\address{Тернопільський національний технічний університет імені Івана Пулюя, \\вул. Руська, 56, 46001 Тернопіль, Україна}

\makeukrtitle

\begin{abstract}
\tolerance=3000%
Використовуючи модель ефективної маси для електрона та моделі діелектричного континууму, отримані аналітичні розв'язки самоузгодженої системи рівнянь Шредінгера-Пуассона.
Для експериментально створеної триямної резонансно-тунельної структури --- каскаду квантового каскадного детектора розроблена квантово-механічна теорія електронних стаціонарних станів, сил осциляторів квантових переходів і метод розрахунку її потенціального профілю.
Для запропонованого методу проведено порівняння з результатами інших методів та результатами експерименту. Отримано добре узгодження між розрахованою величиною детектованої енергії та її експериментальною величиною, які відрізняються не більш, ніж на  $2.5$\%.
\keywords квантовий каскадний детектор, п'єзоелектрична поляризація, спонтанна поляризація, резонансно-тунельна структура,  сила осцилятора

\end{abstract}

\end{document}